\documentclass[journal]{IEEEtran}

\usepackage[usenames]{color}
\usepackage{lscape}
\usepackage{soul}

\newcommand{\AzCom}[1]{}
\newcommand{\AzDel}[1]{}

\newdimen\snellbaselineskip
\newdimen\snellskip
\snellbaselineskip=\baselineskip
\def\srule{\omit\kern.5em\vrule\kern-.5em}
\newbox\bigstrutbox
\setbox\bigstrutbox=\hbox{\vrule height14.5pt depth9.5pt width0pt}
\def\bigstrut{\relax\ifmmode\copy\bigstrutbox\else\unhcopy\bigstrutbox\fi}
\def\middlehrule#1#2{\noalign{\kern-\snellbaselineskip\kern\snellskip}
&\multispan#1\strut\hrulefill
&\omit\hbox to.5em{\hrulefill}\vrule
height \snellskip\kern-.5em&\multispan#2\hrulefill\cr}

\makeatletter
\def\bordermatrix#1{\begingroup \m@th
  \@tempdima 8.75\p@
  \setbox\z@\vbox{%
    \def\cr{\crcr\noalign{\kern2\p@\global\let\cr\endline}}%
    \ialign{$##$\hfil\kern2\p@\kern\@tempdima&\thinspace\hfil$##$\hfil
      &&\quad\hfil$##$\hfil\crcr
      \omit\strut\hfil\crcr\noalign{\kern-\snellbaselineskip}%
      #1\crcr\omit\strut\cr}}%
  \setbox\tw@\vbox{\unvcopy\z@\global\setbox\@ne\lastbox}%
  \setbox\tw@\hbox{\unhbox\@ne\unskip\global\setbox\@ne\lastbox}%
  \setbox\tw@\hbox{$\kern\wd\@ne\kern-\@tempdima\left(\kern-\wd\@ne
    \global\setbox\@ne\vbox{\box\@ne\kern2\p@}%
    \vcenter{\kern-\ht\@ne\unvbox\z@\kern-\snellbaselineskip}\,\right)$}%
  \null\;\vbox{\kern\ht\@ne\box\tw@}\endgroup}
\makeatletter

\makeatletter
\def\bordermatrix#1{\begingroup \m@th
  \@tempdima 8.75\p@
  \setbox\z@\vbox{%
    \def\cr{\crcr\noalign{\kern2\p@\global\let\cr\endline}}%
    \ialign{$##$\hfil\kern2\p@\kern\@tempdima&\thinspace\hfil$##$\hfil
      &&\quad\hfil$##$\hfil\crcr
      \omit\strut\hfil\crcr\noalign{\kern-\snellbaselineskip}%
      #1\crcr\omit\strut\cr}}%
  \setbox\tw@\vbox{\unvcopy\z@\global\setbox\@ne\lastbox}%
  \setbox\tw@\hbox{\unhbox\@ne\unskip\global\setbox\@ne\lastbox}%
  \setbox\tw@\hbox{$\kern\wd\@ne\kern-\@tempdima\left(\kern-\wd\@ne
    \global\setbox\@ne\vbox{\box\@ne\kern2\p@}%
    \vcenter{\kern-\ht\@ne\unvbox\z@\kern-\snellbaselineskip}\,\right)$}%
  \null\;\vbox{\kern\ht\@ne\box\tw@}\endgroup}
\makeatletter

\usepackage{cite}

\ifCLASSINFOpdf
\else
  \usepackage[dvips]{graphicx}
  \DeclareGraphicsExtensions{.eps}
\fi
\usepackage[cmex10]{amsmath}
\usepackage{amssymb}
\usepackage{amsfonts}

\usepackage{url}

\usepackage{psfrag}

\def\Abf{{\mathbf{A}}}
\def\Pbf{{\mathbf{P}}}
\newcommand{\pr}[1]{\Pr \left\{#1\right\}}
\begin{document}
%
\title{Modelling a Decentralized Constraint Satisfaction Solver for Collision-Free Channel Access}
%
%
%

\author{Jaume~Barcelo, 
        Nuria~Garcia, 
        Azadeh~Faridi, \IEEEmembership{Member, IEEE,}
        Simon~Oechsner, 
        and~Boris~Bellalta,~\IEEEmembership{Senior~Member,~IEEE}
\thanks{The authors are with Universitat Pompeu Fabra.
Roc Boronat 138, 08018 Barcelona, Catalunya, Spain.
E-mail: jaume.barcelo@upf.edu
This paper has been submitted to an IEEE journal.
}
}

%
%

\markboth{Journal of \LaTeX\ Class Files,~Vol.~6, No.~1, January~2007}%
{Shell \MakeLowercase{\textit{et al.}}: Bare Demo of IEEEtran.cls for Journals}
%



\maketitle

\begin{abstract}
In this paper, the problem of assigning channel slots to a number of contending stations is modeled as a Constraint Satisfaction Problem (CSP).
A learning MAC protocol that uses deterministic backoffs after successful transmissions is used as a decentralized solver for the CSP.
The convergence process of the solver is modeled by an absorbing Markov chain (MC), and analytical, closed-form expressions for its transition probabilities are derived.
Using these, the expected number of steps required to reach a solution is found.
The analysis is validated by means of simulations and the model is extended to account for the presence of channel errors.
The results are applicable in various resource allocation scenarios in wireless networks.

\end{abstract}

\begin{IEEEkeywords}
Medium Access Control, decentralized constraint satisfaction solver, learning MAC protocol
\end{IEEEkeywords}

%
\IEEEpeerreviewmaketitle

\section{Introduction}
%
%
%
%



\IEEEPARstart{S}{ince} the inception of wireless local area networks (WLANs), random medium access mechanisms have played a key role in arbitrating access to shared channels. The core principles of the medium access control (MAC) that were introduced in the first release of the IEEE 802.11 standard are still valid today \cite{IEEE80211-IEEESTD2007}. The contenders for the channels use carrier sense to avoid interrupting ongoing transmissions. Until recently, slotted time combined with a random backoff have been used to reduce the chances that two stations simultaneously start a transmission. However, it was pointed out recently that the random choice of the backoff value is not necessary after successful transmissions \cite{barcelo2008lba}.
In fact, if all the nodes that have successfully transmitted choose a common deterministic backoff value for their next transmission, the chances of collisions are reduced, since in their next transmission they may only collide with the remaining unsuccessful nodes. Furthermore, under certain conditions, a collision-free operation can be reached and maintained.

The idea of using a deterministic backoff after successful transmissions has been explored in more detail in, e.g., \cite{he2009srb,barcelo2011tcf,fang2011dlm,barcelo2010fcc}.
The goal of this class of protocols is to distributively build a collision-free schedule which can then repeat periodically without further collisions, as long as the network does not change. This is equivalent to the decentralized assignment of stations to slots within one period of the schedule in such a way that no slot is assigned to more than one station. Such an assignment is obtainable if the number of contending stations does not exceed the number of slots in one period of the schedule. We are interested in forecasting the expected number of rounds required to reach a collision-free assignment.

Similar problems can be found in other areas of networking where limited resources need to be distributed among a group of stations. Examples of such resources include channel time slots \cite{barcelo2008lba}, frequency channels \cite{duffy2011dcs}, and code division multiple access scramble codes \cite{checco2012scs}.

In \cite{duffy2011dcs}, a general framework is presented that encompasses problems such as graph coloring, channel assignment to WLANs cells, the search for feasible inter-flow codes in network coding, and the construction of collision-free schedules in CSMA networks.
This framework consists in modeling the resource-assignment problem as a Constraint Satisfaction Problem (CSP) \cite{tsang1993fcs}.
When the nodes that participate in the CSP cannot communicate with one another to solve the problem, a decentralized approach is required.
Decentralized solvers are different from distributed solvers \cite{yokoo2000adc}, which require message exchange among the participating nodes.
The concept of decentralized CSP solvers is very recent and is yet to be explored in depth.
In \cite{duffy2011dcs}, a decentralized solver for this CSP is presented and analyzed, and a bound on the convergence time of the solver is found.

In this work, we focus on analyzing a MAC protocol that uses deterministic backoffs after successful transmissions, however, our results are applicable for many other distributed resource-allocation problems. Similarly to \cite{duffy2011dcs}, we model the channel access problem as a CSP for which the aforementioned protocol serves as a decentralized solver. To calculate the expected number of rounds the solver requires to reach a solution, we model the convergence process using an absorbing Markov chain (MC). The first contribution of this paper is the derivation of closed-form expressions for transition probabilities of the absorbing MC, which are then used to calculate the expected convergence time of the solver. The second contribution is the adaptation of the model to an environment in which errors can occur. 

\section{System Model and the Corresponding CSP}

We consider a wireless network in which channel time is slotted.
The slots are grouped in rounds, and each round contains $B$ consecutive slots.
Our focus is on the distributed assignment of $N$ contending wireless stations to the $B$ channel time slots per round.
In each round, each stations randomly selects one of the $B$ slots in the round.
In each slot a single transmission can be completed and acknowledged.
A station succeeds if the slot it has chosen is not selected by any other station, in which case its transmission is acknowledged in the same slot.
If two or more stations pick the same slot, all stations involved will suffer a collision, in which case no acknowledgment will be received.

\subsection{The Constraint Satisfaction Problem}
A CSP is simply a problem consisting of a set of variables whose value must satisfy a set of constraints. Adopting the notation from \cite{duffy2011dcs}, we consider $N$ variables, \mbox{$\mathbf{x} := (x_1,\dots,x_N)$}, with $x_i \in \mathbf{B}=\left\{1, \dots, B \right\}, \forall i $, and $M$ clauses, $\left\{ \Phi_1(\mathbf{x}),\dots,\Phi_M(\mathbf{x}) \right\}$, that are Boolean functions.
The $M$ clauses represent the constraints and take a value equal to 1 if the constraint is satisfied and 0 otherwise.
An assignment $\mathbf{x}$ is a solution to the problem if all the constraints are satisfied.

For our system model, the variables $\{x_i\}_{i=1}^{N}$ in the corresponding CSP are the slots chosen by the $N$ contending wireless stations from the set of available slots $\mathbf{B}$ in every round. There is one clause per pair of variables evaluating if they have the same value or not. The clause will return 0 if the two participating stations have selected the same slot (i.e., a collision) and returns 1 otherwise.

\subsection{The Decentralized Solver}

Here we describe a protocol that distributively solves the problem posed above, whenever a solution exists. The contention is organized in transmission rounds that contain $B$ transmission slots. In every round, each of the $N$ stations transmit exactly once. A solution to the problem is an assignment in which no slot contains more than one wireless stations. The protocol describes how the stations pick their transmission slots in each round, taking into account that each station is only aware of the outcome of its own transmission in the previous round.

\begin{figure}
  \centering
  \includegraphics[width=2.5in]{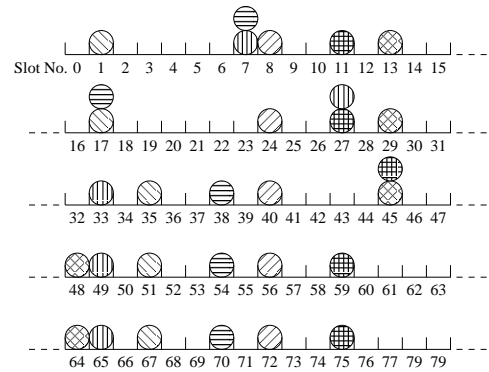}
  \caption{CSMA/ECA contention}
  \label{fig:csma_eca_compact}
\end{figure}

The protocol works as follows. In the first round, each station randomly and independently picks one of the $B$ possible slots in the round. If the transmission is successful, the station will pick exactly the same slot in the next round. Otherwise, it will again pick one of the $B$ slots randomly. This process repeats until all stations successfully transmit in the same round, from which point on, all stations will transmit periodically and no collisions will occur. The operation of this protocol is illustrated in Fig.~\ref{fig:csma_eca_compact} for five consecutive rounds. A collision-free solution is reached in the fourth round, after which the schedule is repeated endlessly without collisions. 

This protocol is a simplified version of a variant of CSMA/CA, called CSMA with enhanced collision avoidance (CSMA/ECA), as detailed in, e.g., \cite{barcelo2008lba,he2009srb,barcelo2009tpc,fang2011dlm}.

This protocol can be viewed as a distributed solver for the CSP defined in the previous subsection. In the first round, the variables $x_i$ are assigned a value in $\{1,\cdots, B \}$ randomly, with probability $1/B$. Then the constraints are evaluated. Those variables that are involved in unsatisfied constraints take a random value again in the next round, while the rest keep the same value as in the previous round. When a solution is reached, all  variables keep the same value. This solver is, in fact, an instance of the parameterized solver in \cite{duffy2011dcs}, with the parameter values set to $a=b=1$. An attractive property of this decentralized solver is that it reaches the solution in finite time, if it exists, and its performance is comparable to the known centralized solvers such as WalkSAT \cite{selman1993lss}.

\section{The Markov Chain Model}
\label{sec:markov_chain}

We are interested in calculating the expected number of rounds required to reach a solution. To this end, we construct a Markov chain to model the behavior of the protocol (or equivalently, the CSP solver). 

By the pigeonhole principle, a solution exists only when $N\leq B$, i.e., when there are at least as many slots in a round as the total number of stations. Considering $N\leq B$ contending stations, the associated MC model has $N+1$ different states, $S_0, \dots, S_N$.
The system is in state $S_d$ if exactly $d$ stations ($0 \leq d \leq N$) were successful in the previous round and, therefore, will deterministically choose their transmission slot in the current round.
From the CSP perspective, this is equivalent to saying that there are exactly $d$ variables that were not involved in any constraint that was not satisfied in the previous round.

We are interested in the computation of the transition probability, $p_{d,\delta}^{B,N}$, from one state $S_d$ to another state $S_\delta$, $0 \leq \delta \leq N$. In other words, $p_{d,\delta}^{B,N}$ is the probability of obtaining $\delta$ successful transmissions given $N$ stations and $B$ slots when $d$ of the stations use a deterministically chosen slot while the remaining $N-d$ stations transmit in a randomly chosen slot.

Note that the considered MC is an absorbing MC, as $p_{N,N}^{B,N}=1$. This is because, once a collision-free schedule is found, the same collision-free schedule is repeated in every subsequent step.
Since the MC is absorbing, the expected number of steps before convergence can be computed if the values of $p_{d,\delta}^{B,N}$ are known \cite{grinstead1997ip}.

\subsection{Calculating the Transition Probabilities}
To calculate the transition probabilities, we number the stations from $1$ to $N$ and define $A_i$ to be the event that station $i$ succeeds, and the set $\mathbf{A}=\{A_i\}_{i=1}^{N}$ to be the collection of all such events.
These events are partially overlapping since more than one station may successfully transmit in the same round. For a given $d$, the transition probability $p_{d,\delta}^{B,N}$ is the probability that exactly $\delta$ out of the $N$ events in $\Abf$ happen. As mentioned before, when $d = N$, the system is in the absorbing state $S_N$, and $p_{N,\delta}^{B,N} = 1$, when $\delta = N$, and is zero otherwise. When $d<N$, this probability can be calculated applying a generalized version of the inclusion-exclusion principle (see, e.g., the theorem in Sec. IV.3 of \cite{feller1968ipt}) as follows:
\begin{equation}
\label{eq:Pij}
p^{B,N}_{d,\delta} = \sum_{j=\delta}^{N} (-1)^{j+\delta}\binom{j}{\delta} S(j),
\end{equation}
where $S(j)$ is given by
\begin{equation}
\label{eq:S_j_def}
S(j) = \sum_{\forall \Abf^{j} \subseteq \Abf} \pr{\bigcap_{A_i \in \Abf^{j}} A_i}.
\end{equation}
Here $\Abf^{j}$ denotes a subset of $\Abf$ that has exactly $j$ elements, i.e., $|\mathbf{A}^{j}|=j$. Therefore, $\pr{\bigcap_{A_i \in \Abf^{j}} A_i}$ is the probability that all the $j$ stations represented in $\Abf^{j}$  successfully transmit. Note that $S(j)$ is a sum of probabilities, but it is not itself a probability.

For a given set of $j$ tagged stations in $\Abf^{j}$, the probability $\pr{\bigcap_{A_i \in \Abf^{j}} A_i}$ depends on $k$, the number of deterministic stations within the $j$ tagged stations. The $j$ tagged stations succeed if the $j-k$ random stations among them choose unoccupied slots, and the remaining $N-d-(j-k)$ untagged random stations choose slots that are different from the ones selected by the $j$ tagged stations. The first event occurs with probability $$\binom{B-d}{j-k} \frac{(j-k)!}{B^{j-k}},$$ and the second with probability $$\left(\frac{B-j}{B} \right)^{N-d-(j-k)}.$$
When $j=N$, we have $k=d$, and therefore, there are no untagged random stations, hence the second probability is 1. Consequently, the probability that all of the $j$ stations of $\Abf^{j}$ succeed, given that $k$ of them are deterministic, after some simplification is
\begin{align}
\label{eq:pAkj}
&\pr{\bigcap_{A_i \in \Abf^{j}} A_i} \\
&~~~~~~~~= \left \{ \begin{array}{lr}
    \displaystyle \frac{(B-d)!(B-j)^{N-d-(j-k)}}{\left(B-d-(j-k)\right)!~B^{N-d}}, & j < N \\&\\
    \displaystyle \frac{(B-d)!}{(B-N)!~B^{N-d}}, & j = N
\end{array}
\right. \nonumber
\end{align}

For any given $j$, there are $\binom{d}{k} \binom{N-d}{j-k}$ sets $\Abf^{j}$ with $k$ deterministic stations. Furthermore, the number of deterministic stations, $k$, among the $j$ tagged stations is bounded by
\begin{align}
\label{eq:k_bound}
\max(0,j+d-N) \leq k \leq \min(d,j),
\end{align}
since in a set of $j$ nodes, there cannot be more deterministic stations than the total number of deterministic stations ($k \leq d$), or more random stations than the total number of random stations ($j-k \leq N-d$).

Using (\ref{eq:S_j_def}), (\ref{eq:pAkj}), and (\ref{eq:k_bound}), $S(j)$ can be calculated as
\begin{align}
\label{eq:S_j}
S(j) = &\sum_{k=\max(0,j+d-N)}^{\min(d,j)}   \binom{d}{k} \binom{N-d}{j-k}\\
 & ~~~~~~~~~~~~\times \frac{(B-d)!(B-j)^{N-d-(j-k)}}{\left(B-d-(j-k)\right)!~B^{N-d}}, ~~~ j < N \nonumber
\end{align}
and for $j=N$,
\begin{align}
\label{eq:S_N}
S(N) = \frac{(B-d)!}{(B-N)!~B^{N-d}}.
\end{align}
Finally, the transition probabilities for $d<N$ can be calculated by replacing $S(j)$ in (\ref{eq:Pij}). 
When $d=0$, i.e., when all the $N$ stations randomly select a slot, this result exactly matches the one obtained in \cite{szpankowski1983asc}.

\subsection{Calculating the Number of Steps until Absorption}

To compute the expected number of rounds needed for the solver to reach a solution, we calculate the expected number of transitions that the MC takes to reach the absorbing state $S_N$ (see, e.g., \cite{grinstead1997ip} for the theory behind this calculation).

Let $\Pbf^{B,N}$ be the transition probability matrix of the MC. This matrix is a square matrix of size $N+1$. If we number the rows and columns of $\Pbf^{B,N}$ starting with zero, the element in row $d$ and column $\delta$ is simply $\left[\Pbf^{B,N}\right]_{d,\delta} = p^{B,N}_{d,\delta}$ as in (\ref{eq:Pij}). In this matrix, rows $0$ to $N-1$ represent transitions from the transient states and row $N$ the transitions from the absorbing state. Therefore, $\Pbf^{B,N}$ has the following general form:
\begin{equation} \label{Eq:Pmat_form}
\Pbf^{B,N} \;= \bordermatrix{
                       &\hbox{TR}  &\omit\hfil &\hbox{ABS}\cr
    \hbox{TR}\bigstrut &\mathbf{Q}_{(N \times N)} &\srule     &\mathbf{c}_{(N \times 1)} \cr
\middlehrule{1}{1}
    \hbox{ABS}\bigstrut&\mathbf{0}_{(1 \times N)} &\srule & 1}
\end{equation}
where $\mathbf{Q}$ is a matrix containing the first $N$ rows and columns of $\Pbf^{B,N}$, from which we calculate the fundamental matrix of the absorbing MC as $\mathbf{N}= (\mathbf{I}_{N \times N}-\mathbf{Q})^{-1}$, where $\mathbf{I}_{N \times N}$ is the $N \times N$ identity matrix.
The expected number of steps to absorption, if the system starts in state $S_0$, is the sum of all the elements in the first row of $\mathbf{N}$.

\subsection{The Markov Chain in the Presence of Channel Errors}

So far we have not considered the possibility that the channel introduces errors. In presence of channel errors, a transmission may be unsuccessful even if it has not suffered a collision. In fact, after an unsuccessful transmission, a wireless station cannot know whether it has suffered a collision or a channel error, and the response of the protocol will be exactly the same, i.e., moving the station back to the random behavior.

In this case, the probability of moving from the state $S_d$ to the state $S_\delta$ is the probability that $i \in \left[ \delta, N\right]$ stations do not collide, but exactly $i-\delta$ of those stations suffer a channel error, i.e.,
\begin{equation}
\label{eq:psiBNepsilon}
p^{B,N,\epsilon}_{d,\delta} = \sum_{i=\delta}^{N} \binom{i}{\delta} \epsilon^{i-\delta}(1-\epsilon)^\delta p^{B,N}_{d,i}.
\end{equation}
where $\epsilon$ is the channel error probability. Note that the resulting MC is no longer absorbing.

\section{Numerical Results}
\label{sec:numerical_results}

\begin{figure}
\centering
\includegraphics[height=6.2cm]{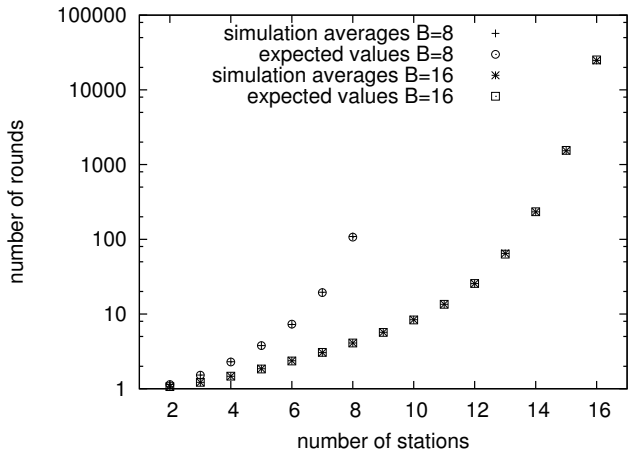}
\caption{The analytically computed expectation is compared to simulation averages. Two different values for $B$ have been considered ($B=8$ and $B=16$) and $N$ takes values from 2 to 16.}
\label{fig:convergence_avg}
\end{figure}

In this section we present simulation results that validate the expressions derived in the previous section.
The number of slots in each round is set to $B=8$ and $B=16$, and the number of contenders $N$ takes values from 2 to 16.
The contenders choose the same slot in the case of successful transmission and a random slot if the transmission is unsuccessful.

The first results are for an ideal channel that does not introduce errors.
We compare the analytically computed expected number of steps to absorption and the average number of steps to reach collision-free operation obtained from 10,000 executions of a custom simulator.\footnote{The two simulators in \emph{C} that we have used and the scripts in \emph{maxima} to compute the expectations derived from the analytical model can be downloaded from \url{https://github.com/jbarcelo/source-paper-mdc} .}
The results are presented in Fig.~\ref{fig:convergence_avg}.

\begin{figure}[t!!!!]
\centering
\includegraphics[height=6.2cm]{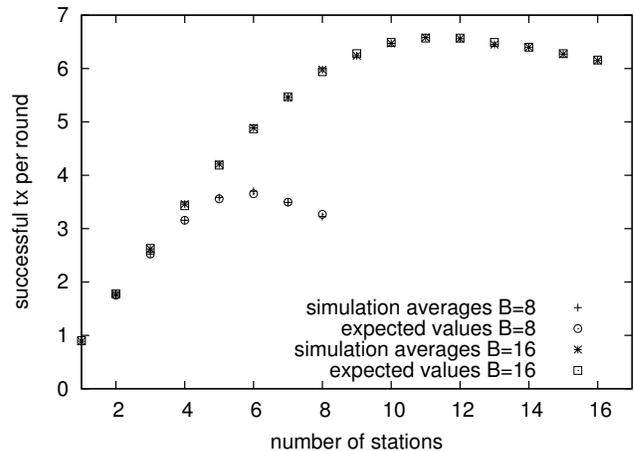}
\caption{The average number of successful transmissions in every round obtained from simulation is compared to the analytically computed expected values.}
\label{fig:successful_tx_per_step}
\end{figure}

To validate the expression in (\ref{eq:psiBNepsilon}) we compute the average number of successful transmissions in each step from the MC and compare it with averages obtained from a simulation of 10,000 rounds.
The results for a channel error probability $\epsilon=0.1$, different numbers of slots ($B$) and different numbers of contenders ($N$) are presented in Fig.~\ref{fig:successful_tx_per_step}.

\section{Conclusion}
\label{sec:conclusion}

We have studied a decentralized CSP solver to assign channel slots to contending stations. With this solver, the system eventually converges to collision-free operation under ideal channel conditions. We have modeled the convergence process as an absorbing MC and have derived closed expressions for the transition probabilities, which are used to compute the expected number of steps required for the system to converge to a solution. We have also considered the presence of channel errors and constructed an MC that accounts for channel errors, and have calculated its transition probabilities. 
The presented results have been validated by means of simulation. The results can be adapted to various scenarios in wireless networks where a finite number of resources need to be distributively assigned to a number of contending stations.

\section*{Acknowledgment}

The authors would like to thank Victor Dalmau and Luis Sanabria for their helpful comments.

\ifCLASSOPTIONcaptionsoff
  \newpage
\fi



\bibliographystyle{IEEEtran}
\bibliography{IEEEabrv,my_bib}
\end{document}